\documentclass[fleqn,twoside]{article}
\usepackage{espcrc2,amssymb,psfig,epsfig}

\usepackage{graphicx}
\usepackage[figuresright]{rotating}

\newcommand{\AmS}{{\protect\the\textfont2
  A\kern-.1667em\lower.5ex\hbox{M}\kern-.125emS}}

\newcommand{\be}{\begin{equation}}
\newcommand{\ee}{\end{equation}}
\newcommand{\bea}{\begin{eqnarray}}
\newcommand{\eea}{\end{eqnarray}}

\def\bsigma{\mbox{\protect\boldmath $\sigma$}}
\newcommand{\reff}[1]{(\ref{#1})}

\title{Spin models on Platonic solids and asymptotic freedom}

\author{Sergio Caracciolo\address{Dipartimento 
         di Fisica, Universit\`a degli Studi di Milano, 
         via Celoria 16, I-20133 Milano, 
         INFN, Sezione di Pisa and NEST-INFM, Italy},
        Andrea Montanari\address{Laboratoire de Physique 
        Th\'eorique de l'Ecole Normale Sup\'erieure,
        Paris, France},
        Andrea Pelissetto\address{Dipartimento di Fisica and INFN --
         Sezione di Roma I, 
         Universit\`a degli Studi di Roma ``La Sapienza", 
         P.le Aldo Moro 2, I-00185 Roma, 
         Italy}
        }
       
\begin{document}

\begin{abstract}
We consider a two-dimensional $\sigma$-model with discrete 
icosahedral/dodecahedral symmetry. We present high-precision 
finite-size numerical results that show that the continuum limit of this model 
is different from the continuum limit of the rotationally invariant 
$O(3)$ $\sigma$-model.
\vspace{1pc}
\end{abstract}

\maketitle

Recently, there has been interest in the critical behavior of 
two-dimensional $\sigma$-models in which the spins take values in some discrete 
subset of the sphere. In particular, two groups 
\cite{PS-98a,PS-98b,PS-01,HN-01a,HN-01b} studied 
the nearest-neighbor $\sigma$-model
\be
H = - \beta \sum_{<ij>} \bsigma_i\cdot \bsigma_j,
\ee
in which the spins have unit length and belong to a Platonic solid, i.e. to a 
tetrahedron, cube, octahedron, icosahedron, or dodecahedron. Several
 quantities 
have been computed, the renormalized two-point function, the 
current-current correlation function, 
the finite-size scaling (FSS) curve for the 
second-moment correlation length,  and the four-point renormalized 
coupling. Surprisingly enough, the results for the icosahedral and the 
dodecahedral model are very close to the $O(3)$ ones, suggesting that  
these three models might have the same continuum limit. 
Patrascioiu and Seiler \cite{PS-98a,PS-98b,PS-01} considered these 
results as evidence for the $O(3)$ $\sigma$-model not
being asymptotically free, since the discrete-symmetry models have 
a finite $\beta$ 
phase transition, which cannot be described in perturbation theory.
However, the overwhelming evidence we have collected in the years in favor 
of asymptotic freedom made Hasenfratz and Niedermayer \cite{HN-01a,HN-01b} 
suggest 
that, may be, the icosahedral and the dodecahedral models have 
an asymptotically-free continuum limit, 
in spite of the fact that the critical point is at a finite value of $\beta$. 

In Ref. \cite{CMP-01} we showed that, by using some standard assumptions, the 
perturbative renormalization-group (RG) approach predicts that the 
suggestion of  Hasenfratz and Niedermayer cannot be true. 
If the continuum limit of the $O(3)$ $\sigma$-model is correctly 
described by the perturbative RG, then any discrete-symmetry model cannot
belong to the same universality class of the $O(3)$ $\sigma$-model. 

The argument goes as follows. Consider the Hamiltonian 
\be
H = - \beta \sum_{<i,j>}\bsigma_i\cdot\bsigma_j - 
     h \sum_i I_n(\bsigma_i), 
\label{Hmisto}
\ee
where $\bsigma_i$ is an $O(3)$ unit spin and $I_n(\bsigma_i)$ 
is a polynomial in $\bsigma_i$ with the following properties:
it has $O(3)$ spin $n$; 
the maxima (or minima) of $I_n(\bsigma_i)$ correspond to a Platonic solid;
it is invariant under the discrete-symmetry group of the solid.
For all Platonic solids,
it can be shown explicitly that such a polynomial exists.
The model \reff{Hmisto} interpolates between the $O(3)$ $(h=0)$ and 
the discrete-symmetry  model $(|h|=+\infty)$. 
Now, with quite standard assumptions, 
one can show that $I_n(\bsigma)$ is a {\em relevant perturbation} of the 
$O(3)$ fixed point. In other words, any arbitrarily small 
perturbation with discrete symmetry 
of the $O(3)$ $\sigma$-model drives the system to a 
different fixed point \cite{CMP-01}. 

\begin{figure}[htb]
\vspace{9pt}
\psfig{file=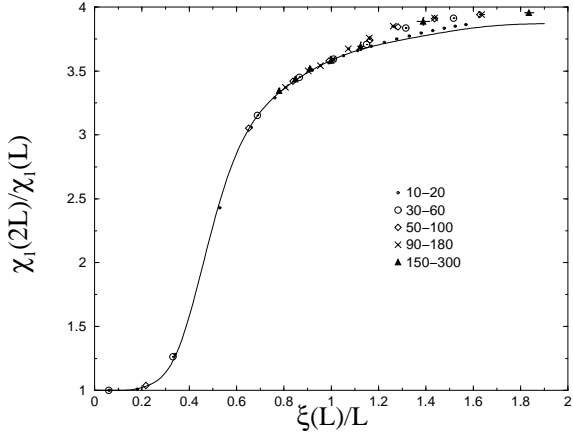,angle=-90,width=1.\linewidth}
\caption{FSS function for the spin-1 susceptibility $\chi_1$. }
\label{chi1}
\end{figure}

\begin{figure}[htb]
\vspace{9pt}
\psfig{file=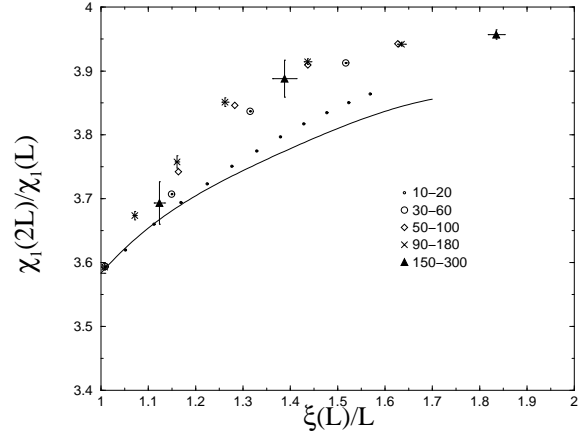,angle=-90,width=1.\linewidth}
\caption{FSS function for the spin-1 susceptibility $\chi_1$. 
Here, we restrict the horizontal range to $1\le x \le 2$.}
\label{chi1bis}
\end{figure}
 
\begin{figure}[htb]
\vspace{9pt}
\psfig{file=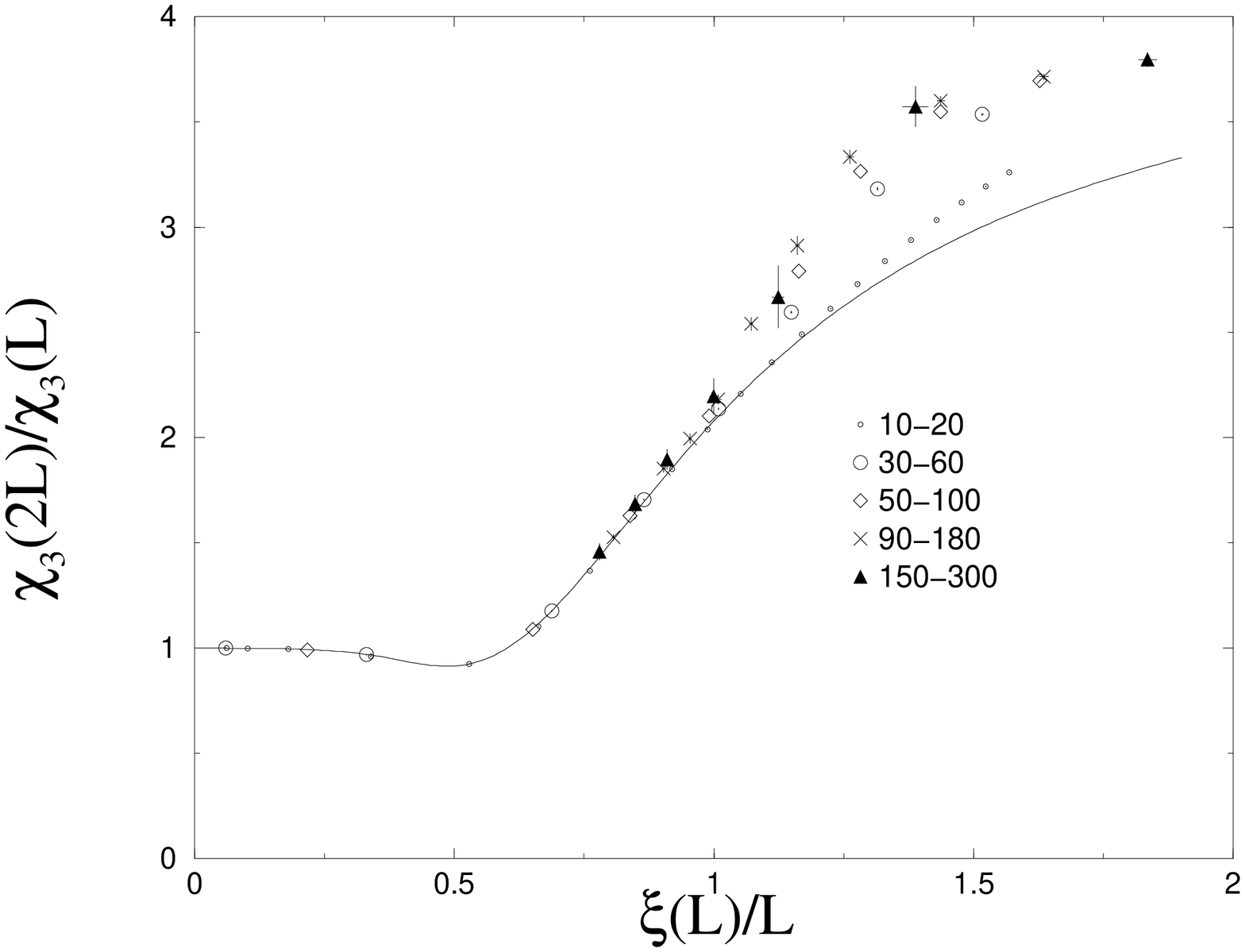,angle=-90,width=1.\linewidth}
\caption{FSS function for the spin-3 susceptibility $\chi_3$.}
\label{chi3}
\end{figure}

\begin{figure}[htb]
\vspace{9pt}
\psfig{file=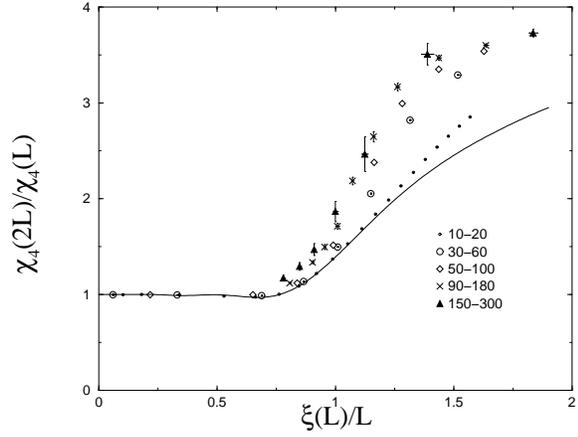,angle=-90,width=1.\linewidth}
\caption{FSS function for the spin-4 susceptibility $\chi_4$.}
\label{chi4}
\end{figure}

This argument together with the numerical results of
Refs.~\cite{PS-98a,PS-98b,PS-01,HN-01a,HN-01b} puts asymptotic freedom 
on a dangerous ground 
since it shows that the conventional scenario is wrong if the icosahedral
or the dodecahedral model have 
the same continuum limit of the $O(3)$ model. We have thus decided 
to extend the previous numerical work and indeed, we have found good 
evidence that the $O(3)$ model and the icosahedral model belong to 
different universality classes: the conventional scenario is saved. 
However, the surprising fact is that these differences appear only 
very near to the critical point, i.e. for $\xi_\infty \gtrsim 10^5$!

In the numerical simulation we have considered the Hamiltonian \reff{Hmisto}
with 
\bea
I_6(\bsigma) &=& \sigma^6_z - 5 \sigma_z^4\left(\sigma_x^2 + \sigma_y^2\right)
   + 5 \sigma_z^2 \left(\sigma_x^2 + \sigma_y^2\right)^2 
\nonumber \\
   && + 2 \sigma_x \sigma_z
     \left(\sigma_x^4 - 10 \sigma_x^2 \sigma_y^2 + \sigma_y^4\right),
\eea
and $h = 0.1$. Such a polynomial is invariant under the rotation group 
of the icosahedron and of the dodecahedron.
We measured the second-moment correlation length 
as defined in Refs.~\cite{CEPS-95,CEFPS-95}, and the spin-$n$ 
susceptibilities
\be
\chi_n = \sum_x \left\langle P_n\left(\sigma_0\cdot\sigma_x\right)\right\rangle,
\ee
where $P_n(x)$ is a Legendre polynomial, for $n=1,3,4$.
Of course, $\chi_1$ is the standard susceptibility.
For each observable ${\cal O}(L,\beta)$, 
we considered the so-called step function,
i.e. the ratio ${\cal O}(2L,\beta)/{\cal O}(L,\beta)$, which,
in the continuum limit should become a universal function of 
$\xi(L,\beta)/L$, i.e.
\be
{{\cal O}(2L,\beta)\over {\cal O}(L,\beta)} = 
   F_{\cal O}\left({\xi(L,\beta)\over L}\right) + O(L^{-\omega},\xi^{-\omega}).
\ee
We measured the step function of these observables in the 
icosahedral theory (i.e. keeping $h=0.1$ fixed) and in the $O(3)$ model,
thereby extending the results of Refs. \cite{CEPS-95,CEFPS-95}.
If the two models have the same continuum limit, the function computed 
for $h=0.1$ and $h=0$ should coincide.

In Figs.~\ref{chi1}, \ref{chi1bis}
we report the numerical results for the spin-one susceptibility.
The continuous line is a fit to the $O(3)$ data, while the points 
refer to the model with $h=0.1$. 
As observed in previous work, there
is indeed very good agreement between the numerical results 
for the two models, but such an agreement disappears for 
$\xi(L)/L \gtrsim 1$, where small discrepancies are observed. 
As it can be seen from Fig.~\ref{chi1bis}, the icosahedral points 
are above the $O(3)$ curve and, more importantly, 
the discrepancy increases with $L$: the points 
with $L=10$--20 are systematically below the points 
with larger values of $L$.

Larger deviations are observed in
Figs.~\ref{chi3} and \ref{chi4}  where we report the FSS curve for 
the spin-3 and spin-4 susceptibilities. Again, the numerical results 
for the icosahedral and the $O(3)$ model agree very nicely up to 
$\xi(L)/L \sim 0.8$-1, but then they indicate that the icosahedral 
FSS curve is steeper than the $O(3)$ one. Again, notice that 
the discrepancy between the results for two models increases with $L$, 
indicating that the observed effect is not due to corrections to scaling, 
i.e. it is not a lattice artifact disappearing in the continuum limit.

In conclusion, the numerical results we have presented 
show that the icosahedral and the $O(3)$
models belong to different universality classes. Note however that 
discrepancies are observed only for $\xi(L)/L \sim 1$, which corresponds 
to very large values of the infinite-volume correlation length. 
For instance, points with  $L=128$ and  $\xi(L)/L \approx 0.94$ 
correspond \cite{CEPS-95} to $\xi_\infty \approx 10^5$.

\end{document}